\documentstyle[12pt,amssymb,amsfonts,amsthm,amssymb,amscd,amstext,
epsfig,amsmath]{article}

\newcommand{\be}{\begin{equation}}
\newcommand{\ee}{\end{equation}}

\newcommand{\bea}{\begin{eqnarray}}
\newcommand{\eea}{\end{eqnarray}}

\def\ba{\begin{array}}
\def\ea{\end{array}}

\begin{document}
\begin{titlepage}
\begin{flushright}
{\bf May 2006} \\ 
DAMTP-06-38\\
hep-th/0605155 \\
\end{flushright}
\begin{centering}
\vspace{.2in}
%
 {\large {\bf Dyonic Giant Magnons}}

\vspace{.3in}

Heng-Yu Chen${}^{1}$, Nick Dorey${}^{1}$ and Keisuke Okamura${}^{2}$ \\
\vspace{.2 in}
${}^{1}$DAMTP, Centre for Mathematical Sciences \\ 
University of Cambridge, Wilberforce Road \\ 
Cambridge CB3 0WA, UK \\
\vspace{.2in}
and \\ 
\vspace{.1 in}
${}^{2}$Department of Physics, Faculty of Science, \\ 
University of Tokyo, Bunkyo-ku,\\
Tokyo 113-0033, Japan.\\ 
\vspace{.4in}

{\bf Abstract} \\

\end{centering}
We study the classical spectrum of string theory on $AdS_{5}\times
S^{5}$ in the Hofman-Maldacena limit. We find a family of classical 
solutions corresponding to Giant Magnons with two independent angular
momenta on $S^{5}$. These solutions are related 
via Pohlmeyer's reduction procedure to 
the charged solitons of the Complex sine-Gordon equation. 
The corresponding string states are dual to BPS boundstates
of many magnons in the spin-chain description of planar
${\cal N}=4$ SUSY Yang-Mills. The exact dispersion relation for these
states is obtained from a purely classical calculation in string
theory.     

\end{titlepage}

\thispagestyle{empty}
\pagebreak
\setcounter{page}{1}
\paragraph{}
The AdS/CFT correspondence predicts that the spectrum of operator
dimensions in planar ${\cal N}=4$ SUSY Yang-Mills and the spectrum of
a free string on $AdS_{5}\times S^{5}$ are the same. Verifying this
prediction by computing the full spectrum is an important unsolved problem. 
For small 't Hooft coupling, $\lambda\ll 1$, the perturbative dimensions of
gauge theory operators can be calculated by diagonalising an integrable spin
chain \cite{MZ,B}. For $\lambda\gg 1$, the string sigma-model, which is also
classically integrable \cite{int}, becomes weakly coupled. The full spectrum
certainly depends in a complicated way on the 't Hooft coupling. 
However the appearance of integrability on both sides of the
correspondence strongly suggests that the problem is solvable and
there has been significant progress in formulating exact 
Bethe ansatz equations which hold 
for all values of the coupling \cite{BDS, AFS, Staud, BS, Dynamic}. 
\paragraph{}
In recent work, Hofman and Maldacena (HM) \cite{HM} 
have identified a particular limit where the problem of determining
the spectrum simplifies considerably. 
In this limit we restrict our attention to Yang-Mills operators with 
large $U(1)$ R-charge $J_{1}$, which therefore also have large scaling
dimensions $\Delta\geq J_{1}$. More precisely, HM consider a limit where
$\Delta$ and $J_{1}$ become infinite with the difference
$\Delta-J_{1}$ and the 't Hooft coupling $\lambda$ held fixed 
(see also \cite{earlier}).    
In this limit both the gauge theory spin chain and the dual string
effectively become infinitely long. The spectrum can then be analysed
in terms of asymptotic states and their scattering. On both sides of
the correspondence the limiting theory is characterised 
by a centrally-extended $SU(2|2)\times SU(2|2)$ supergroup 
which strongly constrains the spectrum and S-matrix \cite{Dynamic}. 
\paragraph{}
The basic asymptotic state carries a conserved momentum $p$, and lies in a 
short multiplet of supersymmetry. States in
this multiplet have different polarisations corresponding to 
transverse fluctuations of the dual string in different directions 
in $AdS_{5}\times S^{5}$. 
The BPS condition essentially determines the 
dispersion relation for all these states 
to be \cite{BDS, Staud, BS, Dynamic} (see also \cite{Dispersion}), 
\begin{eqnarray}
\Delta-J_{1} & = &
 \sqrt{1+\frac{\lambda}{\pi^{2}}\sin^{2}\left(\frac{p}{2}\right)}\,.
\label{disp1}
\end{eqnarray}
In the spin chain description, this multiplet corresponds to an elementary
excitation of the ferromagnetic vacuum known as a magnon. 
The dual state in semiclassical string theory was identified in \cite{HM}.  
It corresponds to a localised classical soliton which propagates 
on an infinite string moving on an ${\mathbb{R}}\times 
S^2$ subspace of $AdS_{5}\times S^{5}$. 
The conserved magnon momentum $p$ 
corresponds to a certain geometrical angle in the target space
explaining the periodic momentum dependence appearing in
(\ref{disp1}). Following \cite{HM}, we will 
refer to this classical string configuration as a Giant Magnon.
\paragraph{}
In addition to the elementary magnon, the asymptotic spectrum of the
spin chain also contains an infinite tower of boundstates
\cite{Dorey}. Magnons with polarisations in an $SU(2)$ subsector 
carry a second conserved $U(1)$ R-charge, denoted $J_{2}$, and form 
boundstates with the exact dispersion relation,   
\begin{eqnarray}
\Delta-J_{1} & = &
 \sqrt{J_{2}^{2}+\frac{\lambda}{\pi^{2}}\sin^{2}\left(\frac{p}{2}\right)}\,.
\label{disp3}
\end{eqnarray}     
The elementary magnon in this subsector 
has charge $J_{2}=1$ and states with $J_{2}=Q$
correspond to $Q$-magnon boundstates. 
These states should exist for all integer values of $J_{2}$ and for
all values of the 't Hooft coupling \cite{Dorey}. In particular we are free to
consider states where $J_{2}\sim \sqrt{\lambda}$. For such states 
the dispersion relation (\ref{disp3}) has the appropriate scaling for
a classical string carrying a second large classical angular
momentum $J_{2}$. In this Letter we will identify the corresponding
classical solutions of the worldsheet theory and determine some of
their properties. In particular, we will reproduce the exact BPS dispersion 
relation (\ref{disp3}) from a purely classical calculation 
in string theory.  In the case of 
boundstates with momentum $p=\pi$, the relevant configuration was
already obtained in \cite{Dorey} as a limit of the two-spin folded
string solution of \cite{FT,AFRT}. Here we will obtain the general
solution for arbitrary momentum $p$ and R-charge $J_{2}$. Both the $p=\pi$ 
configuration of \cite{Dorey} and the original single-charge 
Giant Magnon of HM will emerge as special cases. We also briefly
discuss semiclassical quantisation of these objects which simply has
the effect of restricting the R-charge $J_{2}$ to integer values.        
\paragraph{}
The minimal string solutions carrying two independent angular momenta,
$J_{1}$ and $J_{2}$ correspond to strings moving on an 
${\mathbb{R}}\times S^{3}$ subspace of $AdS_{5}\times S^{5}$. 
In static gauge, the string
equations of motion are essentially those of a bosonic $O(4)$ 
$\sigma$-model supplemented by the Virasoro constraints. An efficient way 
to find the relevant classical solutions exploits the equivalence 
of this system to the Complex sine-Gordon (CsG) equation 
discovered many years ago
by Pohlmeyer \cite{Pohl}\footnote{The corresponding reduction of the $O(3)$
sigma model to the ordinary sine-Gordon equation \cite{Pohl} 
was discussed in \cite{HM} (see also \cite{Mik}).}. The CsG equation 
is completely integrable and has a family of soliton 
solutions \cite{CSG,DH,BT}. In addition to a
conserved momentum the soliton also carries an additional conserved
charge associated with rotations in an internal space. 
Each CsG soliton corresponds to a classical
solution for the string. The problem of reconstructing the
corresponding string
motion, while still non-trivial, involves solving linear differential
equations only. We construct the 
two-parameter family of string solutions corresponding to a single CsG
soliton and show that they have all the expected properties of Giant Magnons. 
In particular they carry
non-zero $J_{2}$ and obey the BPS dispersion relation
(\ref{disp3}). It is quite striking that we obtain the {\em exact} BPS
formula, for all values of $J_{2}$, from a classical calculation. 
This situation seems to be very analogous to that of BPS-saturated
Julia-Zee dyons in ${\cal N}=4$ SUSY Yang-Mills \cite{JZ,TW}. These
objects also have a classical BPS mass formula which turns out to
be exact. It seems appropriate to call our new two-charge 
configurations Dyonic Giant Magnons. 
\paragraph{}
Multi-soliton solutions of the CsG equation are also available in the
literature \cite{CSG,DH,BT}. In the classical theory these objects undergo 
factorised scattering with a known time-delay. This is precisely the 
information required to calculate the semiclassical approximation to
the S-matrix and we hope to return to this in the near future. 
The rest of the paper is
organised as follows. We begin by discussing strings on 
${\mathbb{R}}\times S^{3}$ in the HM limit and review the original Giant
Magnon solution of \cite{HM} in this context. We then explain the
classical equivalence of the string equations to the complex
sine-Gordon equation. Finally, we construct the required string solutions
from the CsG solitons and determine their properties. 
       
\begin{figure}[t]
\begin{center}
\hspace{-.0cm}\includegraphics[scale=0.75]{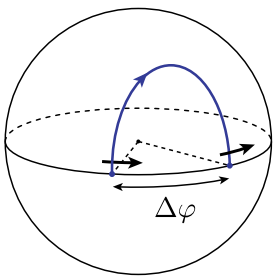}
\caption{\small A Giant Magnon solution.
The endpoints of the string move on the equator 
$\theta=\pi/2$ at the speed of light.
The magnon momentum is given by $p=\Delta\varphi$, where
$\Delta\varphi$ is the angular distance between 
two endpoints of the string.}
\label{fig:giant-magnon}
\end{center}
\end{figure}

\paragraph{}
We will begin by focusing on closed bosonic strings moving
on an ${\mathbb{R}}\times S^{3}$ subspace of $AdS_{5}\times S^{5}$. 
The worldsheet coordinates are denoted 
$\sigma\sim \sigma+2\pi$ and $-\infty<\tau<\infty$ while those on the
target space are, 
\begin{equation}
X_{0}(\sigma,\tau)\,,\;\;\;\vec{X}(\sigma,\tau)=
(X_{1},X_{2},X_{3},X_{4})~~{\mathrm{with}}~~|\vec{X}|^{2}=1 
\label{worldsheetfields}\,.
\end{equation}  
We make the static gauge choice $X_{0}(\sigma,\tau)=\kappa\tau$, 
so that the string energy is given as
$\Delta=\sqrt{\lambda}\,\kappa$. We define a $U(1)_{1}\times U(1)_{2}$ 
Cartan subgroup of the $SU(2)_{L}\times
SU(2)_{R}$ isometry group of the target space 
under which the complex coordinates $Z_{1}=X_{1}+iX_{2}$ and
$Z_{2}=X_{3}+iX_{4}$ have charges $(1,0)$ and $(0,1)$
respectively. String states carry the corresponding conserved Noether 
charges, 
\begin{equation}
J_{1}=\frac{\sqrt{\lambda}}{2\pi}\int^{2\pi}_{0}d\sigma\,
{\mathrm{Im}}[\bar{Z}_{1}
\partial_{\tau}Z_{1}]\label{J1}\,,
\end{equation}
\begin{equation}
J_{2}=\frac{\sqrt{\lambda}}{2\pi}\int^{2\pi}_{0}d\sigma\,
{\mathrm{Im}}[\bar{Z}_{2}\partial_{\tau}Z_{2}]\label{J2}\, ,
\end{equation} 
which can be thought of as 
angular momenta in two orthogonal planes within $S^{3}$.
\paragraph{}
We will now describe the HM limit \cite{HM} where one of these
angular momenta, say $J_{1}$, becomes large 
while the other, $J_{2}$, is held fixed.  
The specific limit we consider is, 
\begin{eqnarray}
&&J_{1}\to \infty\,,\;\;\;\;\;\Delta\to \infty\nonumber \\
&&\Delta-J_{1}={\mathrm{fixed}}\,,\;\;\;\;\lambda={\mathrm{fixed}}\,,\;\;\;\;\;
J_{2}={\mathrm{fixed}}\label{fixed}\,.
\end{eqnarray}
The fact that $\lambda$ is held fixed allows us to interpolate
between the regimes of small and large $\lambda$, 
where perturbative gauge theory and semiclassical string theory
respectively are valid. 
It is convenient to implement the HM limit for the string by
defining the following rescaled worldsheet coordinates, 
$(x,t)\equiv(\kappa\sigma,\kappa\tau)$, which are held fixed as 
$\kappa=\Delta/\sqrt{\lambda}\to\infty$. 
Under this rescaling, the interval $-\pi \le \sigma \le \pi$
corresponding to the closed string is mapped to the real line 
$-\infty\le x \le \infty$ with the point 
$\sigma=\pm\pi$ mapped to $x=\pm\infty$. 
\paragraph{}
As always, a 
consistent closed string configurations always involve 
at least two magnons with zero total momentum. As explained in
\cite{HM}, magnon momentum is associated with a certain geometrical
angle in the target space. The condition that the total momentum
vanishes (modulo $2\pi$) is then enforced by the closed string
boundary condition.   
However, after 
the above rescaling, 
closed string boundary condition and thus the vanishing of the total
momentum can actually be relaxed. 
This allows us to focus on a single worldsheet excitation or magnon
carrying non-zero momentum $p$. 
This makes sense because the additional magnon with momentum $-p$ required 
to make the configuration consistent can be hidden at the point 
$x=\infty$.
\paragraph{} 
The conserved charges of the system which remain finite in the HM limit
are given as, 
\begin{equation}
\Delta-J_{1}=\frac{\sqrt{\lambda}}{2\pi}\int^{+\infty}_{-\infty}dx\,
(1-{\mathrm{Im}}
[\bar{Z}_{1}\partial_{t}Z_{1}])\label{E-J1}\,,
\end{equation}
\begin{equation}
J_{2}=\frac{\sqrt{\lambda}}{2\pi}\int^{\infty}_{-\infty}dx\,{\mathrm{Im}}
[\bar{Z}_{2}\partial_{t}Z_{2}]\label{J2*}\,.
\end{equation}  
It is important to note that 
these quantities will not necessarily be equal to their counter-parts 
(\ref{J1},\ref{J2}) 
computed in the original worldsheet coordinates. In general, the latter may
include an additional contribution coming from a neighbourhood of the 
point $\sigma=\pm\pi$ which is mapped to $x=\infty$. 
As in the above discussion of worldsheet momentum, the extra contribution 
reflects the presence of additional magnons at infinity.       
\paragraph{}
The equation of the motion for the target space coordinate $\vec{X}(x,t)$
can be written in terms of light-cone coordinates $x_{\pm}=(t\pm x)/2$
as 
\begin{equation}
\partial_{+}\partial_{-}\vec{X}+(\partial_{+}\vec{X}\cdot\partial_{-}\vec{X})
\vec{X}=0
\label{eomX}\,.
\end{equation}
A physical string solution must also satisfy the
Virasoro constraints. In terms of the rescaled coordinates these
become,  
\begin{equation}
\partial_{+}\vec{X}\cdot\partial_{+}\vec{X}=
\partial_{-}\vec{X}\cdot\partial_{-}\vec{X}=1
\label{virasoro}\,.
\end{equation}
\paragraph{}
Before discussing the general two-charge case, we will review the
simpler situation where $J_{2}=0$ which leads to the basic Giant
Magnon solution of \cite{HM}. This corresponds to restricting our
attention to strings which are constrained to lie on 
an $S^{2}$ submanifold of
$S^{3}$. In terms of the worldsheet fields introduced 
above, we can implement this by setting $X_{4}(x,t)=0$ or equivalently
demanding that $Z_{2}(x,t)$ is real. In this case the complex worldsheet
fields $Z_{1}$ and $Z_{2}$ can be written as, 
\begin{equation}
Z_{1}(x,t)=\sin\theta\exp(i\varphi)\,,\;\;\;\;\;Z_{2}(x,t)=
\cos\theta\label{hmz1z2}\,,
\end{equation}
where $\theta$ and $\varphi$ are the polar and azimuthal angles on
the two-sphere respectively. In these coordinates the angular momentum
$J_{1}$ generates shifts of the azimuthal angle $\varphi$. 
The required solution should have
infinite energy $\Delta$ and angular momentum $J_{1}$ with a finite
difference $\Delta-J_{1}$. Such a configuration can be obtained by
considering an open string\footnote{Recall that in the HM limit we
have relaxed the closed string boundary condition. To obtain a
consistent closed string configuration we should add a second magnon
at infinity in the coordinate $x$. In spacetime this corresponds to
adding a second open string to form a folded closed string.} 
with both endpoints moving 
on the equator $\theta=\pi/2$ at the 
speed of light. The string theory quantity corresponding to the magnon
momentum $p$ is exactly given by the 
angular separation $\Delta\varphi$ between these two endpoints of the 
string (see Figure \ref{fig:giant-magnon}). 
The Giant Magnon solution therefore has the following boundary
conditions for $Z_{1}$ and $Z_{2}$, 
\begin{equation}
Z_{1}\to \exp\left(it\pm i\frac{p}{2}\right)\,,\;\;\;\;\; Z_{2}\to 0,
~~{\mathrm{as}}~~x\to\pm\infty\label{bc}\,.
\end{equation}

In the following we will show that the unique solution with the
required properties which satisfies (\ref{bc}) is
\begin{equation}
Z_{1}  =  \left[\sin\left(\frac{p}{2}\right)\tanh(Y)-
i\cos\left(\frac{p}{2}\right)\right]\,\exp(i t)\,,\;\;\;\;\; 
Z_{2}  =
\frac{\sin\left(\frac{p}{2}\right)}{\cosh(Y)}\label{hm}\,,
\end{equation}
where
\begin{equation}
Y=\frac{x-\cos\left(\frac{p}{2}\right)t}
{\sin\left(\frac{p}{2} \right)}\,. 
\end{equation}
This solution is equivalent to the one given as Eqn (2.16) in
\cite{HM}. We will rederive it as a special case of the
more general solution presented below. One
may check that while the energy $\Delta$ and the angular momentum
$J_{1}$ of the 
solution diverge, the combination $\Delta-J_{1}$ remains
finite and is given by, 
\begin{equation}
\Delta-J_{1}=\frac{\sqrt{\lambda}}{\pi}\left|\sin\left(\frac{p}{2}\right)
\right|\label{hmE-J1}
\end{equation}
in agreement with the large-$\lambda$ limit of (\ref{disp1}). 
 Moreover the solution (\ref{hm}) carries only one non-vanishing
angular momentum, having $J_{2}=0$. 
\paragraph{}
To solve the string equations of motion (\ref{eomX}) in the general
case, together with the
Virasoro conditions (\ref{virasoro}), we will exploit the 
equivalence of this system with the CsG
equation. Following \cite{Pohl}, we will begin by
identifying the the $SO(4)$ 
invariant combinations of the worldsheet fields $\vec{X}$ and their
derivatives. As the first derivatives $\partial_{\pm}\vec{X}$ are unit 
vectors, we can define a real scalar field $\phi(x,t)$ via the 
relation,    
\begin{equation}
\cos\phi=\partial_{+}\vec{X}\cdot\partial_{-}\vec{X}\label{cosphi}\,.
\end{equation}
Taking into account the constraint $|\vec{X}|^{2}=1$, we see that 
there are no other independent $SO(4)$ invariant quantities that can be
constructed out of the fields and their first derivatives. At the
level of second derivatives we can construct two additional
invariants;     
\begin{equation}
u\sin\phi =\partial_{+}^{2}\vec{X}\cdot \vec{K}\,,\;\;\; 
v\sin\phi =\partial_{-}^{2}\vec{X}\cdot\vec{K}\label{uv}\,,
\end{equation}
where the components of vector $\vec{K}$ are given by 
$K_{i}=\epsilon_{ijkl}X_{j}\partial_{+}X_{k}\partial_{-}X_{l}$.
The connection with CsG model arises from the equations of motion for
$u$, $v$ and $\phi$ derived in \cite{Pohl}. 
In fact the resulting equations imply 
that $u$ and $v$ are not independent and can be
eliminated in favour of a new field $\chi(x,t)$ as, 
\begin{equation}
u=\partial_{+}\chi\tan\left(\frac{\phi}{2}\right)\,,\;\;\;\;\;
v=-\partial_{-}\chi\tan\left(\frac{\phi}{2}\right)\label{newuv}\,.
\end{equation}
The equations of motion for $\chi$ and $\phi$ can then be written as
\begin{eqnarray}
&&\partial_{+}\partial_{-}\phi+\sin\phi-\frac{\tan^{2}
\left(\frac{\phi}{2}\right)}{\sin\phi}
\partial_{+}\chi\partial_{-}\chi=0
\label{eomphi}\,,\\
&&\partial_{+}\partial_{-}\chi+\frac{1}{\sin\phi} 
(\partial_{+}\phi\partial_{-}\chi+
\partial_{-}\phi\partial_{+}\chi)=0\label{eomchi}\,.
\end{eqnarray}
In the special case of constant $\chi$ they reduce to the usual
sine-Gordon equation for $\phi(x,t)$. 
Finally we can combine the real fields $\phi$ and $\chi$ to form a  
complex field
$\psi=\sin\left(\phi/2\right)\exp(i\chi/2)$, which
obeys the equation,
\begin{equation} 
\partial_{+}\partial_{-}\psi+
\psi^{*}
\frac{\partial_{+}\psi\partial_{-}\psi}{1-|\psi|^{2}}+\psi(1-|\psi|^{2})=0\,.
\label{csg}
\end{equation}
\paragraph{}
Equation (\ref{csg}) is known as the Complex sine-Gordon equation. Like the
ordinary sG equation, it is completely integrable and has 
localised soliton solutions which undergo factorised scattering. 
The CsG equation is invariant under a global rotation of the phase of
the complex field: $\psi\rightarrow \exp(i\nu)\psi$, 
$\psi^{*}\rightarrow \exp(-i\nu)\psi^{*}$. In addition to momentum and
energy, CsG solitons carry the corresponding conserved $U(1)$ 
Noether charge\footnote{Note that there is no simple relation 
between the $U(1)$ 
charge of the CsG soliton and the string angular momentum $J_{2}$.}.  
The most general one soliton solution to (\ref{csg}) is given by (see
eg \cite{DH}), 
\begin{eqnarray}
\psi_{\rm 1\mbox{-}soliton} &= & e^{i\mu}
\frac{\cos(\alpha) \exp(i\sin(\alpha)
  T)}{\cosh(\cos(\alpha)(X-X_{0}))} 
\label{onesol}
\end{eqnarray}
with
\begin{eqnarray}
X=\cosh(\theta)x-\sinh(\theta) t\,, & \;\;\;\;\;
&  T=\cosh(\theta)t-\sinh(\theta)x\label{XandT}\,.
\end{eqnarray} 
The constant phase $\mu$ is irrelevant for our purposes as only the
derivatives of the field $\chi$ affect the corresponding string
solution. The parameter $X_{0}$ can be absorbed by a constant 
translation of the world-sheet coordinate $x$ and we will set it to
zero. The two remaining parameters of the solution 
are the rapidity $\theta$ of the soliton and an additional  
real number $\alpha$ which determines the $U(1)$ charge carried by the
soliton.   
\paragraph{}
Taking the limit $\alpha\rightarrow 0$, the field $\phi$ corresponding
to the one-soliton solution (\ref{onesol}) reduces to the kink
solution of the ordinary sG equation. As it is the only known solution of
the CsG equation with this property, it is the unique candidate for the  
dyonic Giant Magnon solution we seek. It remains to reconstruct the
corresponding configuration of the string worldsheet fields $\vec{X}$ 
(or equivalently $Z_{1}$ and $Z_{2}$) corresponding to (\ref{onesol}) 
for general values of the rapidity $\theta$ and rotation parameter $\alpha$.   
\label{stationary} 
\paragraph{}
In this case we have, 
\begin{equation}
\partial_{+}\vec{X}\cdot\partial_{-}\vec{X}=\cos(\phi)=1-
\frac{2\cos^{2}(\alpha)}{\cosh^{2}\left(\cos(\alpha)X\right)}\,.
\label{sc1}
\end{equation}
Hence the complex coordinates $Z_{1}$ and $Z_{2}$ must both solve 
the linear equation, 
\begin{eqnarray}
\frac{\partial^{2}Z}{\partial t^{2}}-
\frac{\partial^{2}Z}{\partial x^{2}}+\left[1-
\frac{2\cos^{2}(\alpha)}{\cosh^{2}\left(\cos(\alpha)X\right)}\right]Z & =
& 0 
\label{xt1}
\end{eqnarray}
where, as above $X=\cosh(\theta)x-\sinh(\theta)t$ and 
we impose the boundary conditions appropriate for a Giant
Magnon with momentum $p$,
\begin{equation}
Z_{1}\to \exp\left(it\pm i\frac{p}{2}\right)\,\;\;\;\;\; Z_{2}\to 0,
~~{\mathrm{as}}~~x\to\pm\infty\label{bc2}\,.
\end{equation} 
As always the two complex fields obey the 
constraint $|Z_{1}|^{2}+|Z_{2}|^{2}=1$. 
We will find unique solutions of the linear equation (\ref{xt1})
obeying these conditions and then, for self-consistency, check that
they correctly reproduce (\ref{sc1}).   
\paragraph{}
It is convenient to express the solution of (\ref{xt1}) 
in terms of the boosted coordinates $X$ and $T$. In terms of these
variables $Z=Z[X,T]$ obeys, 
\begin{eqnarray}
\frac{\partial^{2}Z}{\partial T^{2}}-
\frac{\partial^{2}Z}{\partial X^{2}}+\left[1-
\frac{2\cos^{2}(\alpha)}{\cosh^{2}\left(\cos(\alpha)X\right)}\right]Z & =
& 0\,. 
\label{xt2}
\end{eqnarray} 
The problem now has the form of a Klein-Gordon equation describing the
scattering of a relativistic particle in one spatial dimension 
incident on a static potential well.    
As usual the general solution of this equation can be written as a linear
combination of ``stationary states'' of the form, 
\begin{equation} 
Z_{\omega}=F_{\omega}(X)\exp(i\omega T)\,.
\end{equation} 
Rescaling the variables according to, 
\begin{eqnarray} 
\xi =\cos(\alpha)X \,,\,\,\,\,\, & f(\xi)=F_{\omega}(X) \,,\,\,\,\,\, & 
\varepsilon= 
\frac{\sqrt{\omega^{2}-1}}{\cos(\alpha)}\,,
\end{eqnarray}
we find that the function $f(\xi)$ obeys the equation, 
\begin{equation}
-\frac{d^{2}f}{d\xi^{2}}-\frac{2}{\cosh^{2}(\xi)}\, f=\varepsilon^{2}f\,.
\label{RM}
\end{equation}
\paragraph{}
Equation (\ref{RM}) coincides with the time-independent
Schr\"{o}dinger equation for a particle in (a special case of) 
the Rosen-Morse potential 
\cite{RM}, 
\begin{equation}
V(\xi)= \frac{-2}{\cosh^{2}(\xi)}\,.
\label{RM2}
\end{equation}
The exact spectrum of this problem is known (see {\it{e.g.}}\! \cite{LL}). There is a single normalisable boundstate with 
energy $\varepsilon^{2}=-1$ and wavefunction, 
\begin{equation}
f_{-1}(\xi)=\frac{1}{\cosh(\xi)}
\label{bs}  
\end{equation}
and a continuum of scattering states with $\varepsilon^{2}=k^{2}$ for 
$k>0$ and wavefunctions, 
\begin{equation}
f_{k^{2}}(\xi)=\exp(ik\xi)\left(\tanh(\xi)-ik\right)
\label{scat}
\end{equation}
with asymptotics, 
\begin{equation}
f_{k^{2}}(\xi)\rightarrow \,\,\exp\left(ik\xi\pm i\frac{\delta}{2}\right)
\label{asymp}
\end{equation}
where the scattering phase-shift is given as
$\delta=2\tan^{-1}(1/k)$. 
\paragraph{}
The general solution to the original linear
equation (\ref{xt1}) can be constructed as a linear combination of
these boundstate and scattering wavefunctions. The particular
solutions corresponding to the worldsheet fields $Z_{1}$ and $Z_{2}$ 
are singled out by the boundary conditions (\ref{bc2}). 
In particular, the boundary condition (\ref{bc2}) can only be matched 
by a solution corresponding to a single scattering mode $f_{k^{2}}(\xi)$; 
\begin{equation} 
Z_{1}=c_{1}f_{k^{2}}\left(\cos(\alpha)X\right)\,\exp(i \omega_{k^{2}} T)
\end{equation}         
\paragraph{}
where $\omega_{k^{2}}=\sqrt{k^{2}\cos^{2}(\alpha)+1}$. We find that (\ref{bc2})
is obeyed provided we set, 
\begin{equation}
k=\frac{\sinh(\theta)}{\cos(\alpha)} 
\label{k}
\end{equation}
which yields the magnon momentum $p=\delta=2\tan^{-1}(1/k)$. The
boundary condition (\ref{bc2}) dictates that $Z_{2}$ decays at left
and right infinity. This is only possible if we identify it with the
solution corresponding to the unique normalisable boundstate of the
potential (\ref{RM}), 
\begin{equation}  
Z_{2}=c_{2}f_{-1}(\cos(\alpha)X)\exp(i\omega_{-1}T)
\end{equation} 
with $\omega_{-1}=\sin(\alpha)$. Without loss of generality we can choose the
constants $c_{1}$ and $c_{2}$ to be real. The 
condition $|Z_{1}|^{2}+|Z_{2}|^{2}=1$ then yields, 
\begin{equation}  
c_{1}=c_{2}=\frac{1}{\sqrt{1+k^{2}}}\,.
\end{equation}
\paragraph{}
To summarise the above discussion the resulting string solution is, 
\begin{eqnarray}
Z_{1} & = &
\frac{1}{\sqrt{1+k^{2}}}\,\,\left(\tanh\left[\cos(\alpha)X\right]
-ik\right)\,\, \exp(it)\,, \nonumber \\ 
Z_{2} & = &
\frac{1}{\sqrt{1+k^{2}}}\,\,\frac{1}{\cosh\left[\cos(\alpha)X\right]} 
\exp\left(i\sin(\alpha)T\right)\,,
\label{soln}
\end{eqnarray}
where $X$, $T$ and $k$ are defined in (\ref{XandT}) and (\ref{k})
above. One may easily check that
this solution, in addition to obeying the string equation of motion (\ref{xt1})
and boundary conditions (\ref{bc2}), obeys the Virasoro
constraints and satisfies the self-consistency condition (\ref{sc1}). 
It also reduces to the Hofman-Maldacena solution (\ref{hm}) 
in the non-rotating case $\alpha=0$. Setting $p=\pi$, we obtain 
one-half\footnote{See footnote below Eqn (\ref{hmz1z2}).} 
of the folded string configuration discussed in \cite{Dorey}.  
\paragraph{}
The solution (\ref{soln}) depends on two parameters: $k$ and
$\alpha$. We can now evaluate the conserved charges
(\ref{E-J1}) and (\ref{J2*}) as a function of these parameters, 
\begin{eqnarray}
\Delta-J_{1} & = & \frac{\sqrt{\lambda}}{\pi}\, \frac{1}{1+k^{2}}\,\, 
\frac{\sqrt{1+k^{2}\cos^{2}(\alpha)}}{\cos(\alpha)}\,, \nonumber \\ 
J_{2} & = & \frac{\sqrt{\lambda}}{\pi}\, \frac{1}{1+k^{2}}\,\,
\tan(\alpha)\,. 
\label{param}
\end{eqnarray}  
As above the magnon momentum is identified as $p=2\tan^{-1}(1/k)$. 
Eliminating $k$ and $\alpha$ we obtain the dispersion relation, 
\begin{eqnarray}
\Delta-J_{1} &= & \sqrt{J_{2}^{2}+\frac{\lambda}{\pi^{2}}\sin^{2}
\left(\frac{p}{2}\right)}\,,
\label{res}
\end{eqnarray}
which agrees precisely with the BPS dispersion relation (\ref{disp3}) for
the magnon boundstates obtained in \cite{Dorey}.  
\paragraph{}
The time dependence of the solution (\ref{soln}) is also of interest. 
As in the orginal HM solution the 
constant phase rotation of $Z_{1}$ with exponent $it$
ensures that the endpoints of the string move on an equator of the
three-sphere at the speed of light. We can remove this dependence by
changing coordinates from $Z_{1}$ to $\widetilde{Z}_{1}=\exp(-it)Z_{1}$. 
In the new frame, the string configuration 
depends periodically on time through 
the $t$-dependence of $Z_{2}$. The period, ${\cal T}$, 
for this motion is the time 
for the solution to come back to itself up to a translation of
the worldsheet coordinate $x$. From (\ref{soln}) we find, 
\begin{equation}
{\cal T}=2\pi \frac{\cosh(\theta)}{\sin(\alpha)}\,.
\label{t}
\end{equation}
\paragraph{}  
As we have a periodic classical solution it is natural to define a
corresponding action variable. A leading-order semiclassical
quantization can then be performed by restricting the action variable
to integral values according to the Bohr-Sommerfeld condition. 
Following \cite{HM}, the action
variable $I$ is defined by the equation, 
\begin{equation} 
dI=\frac{{\cal T}}{2\pi}\, \left. d(\Delta-J_{1})\right|_{p} \,.
\end{equation} 
where the subscript $p$ indicates that the differential is taken with
fixed $p$. Using (\ref{param}), (\ref{res}) and (\ref{t}) we obtain simply 
$dI=dJ_{2}$ which is consistent with the identification $I=J_{2}$.  
This is very natural as we expect the angular momentum $J_{2}$ to be 
integer valued in the quantum theory. It is also consistent with the
semiclassical quantization of finite-gap solutions discussed in
\cite{DV} where the action variables correspond to the filling
fractions. These quantities are simply the number of units
of $J_{2}$ carried by each worldsheet excitation.     
\paragraph{}
HYC is supported by a Benefactors' scholarship from St.\,John's
College, Cambridge. ND is supported by a PPARC Senior Fellowship.

\end{document}